\documentclass[12pt]{article}
\usepackage{amssymb}

\begin{document}
%
%
\def\del{\partial}
\def\Ds{D\!\!\!\! /}
\def\Dsbar{\bar{D\!\!\!\! /}}
\def\qu{\mathit{Q}}
\def\qubar{\bar\mathit{Q}}
\def\qbar{\bar\mathcal{Q}}
\def\q{\mathcal{Q}}
\def\w{\mathcal{W}}
\def\S{\mathcal{S}}
\def\P{\mathcal{P}}
\def\D{\mathcal{D}}
\def\B{\mathcal{B}}
\def\X{\mathcal{X}}
\def\wbar{\bar{\mathcal{W}}}
\def\vD{\vec{D}}
\def\vt{\vec{\tau}}
\def\kok{\sqrt{2}}
\def\yarim{{{1}\over{2}}}
\def\lb{\lbrace}
\def\rb{\rbrace}
\def\a{\alpha}
\def\b{\beta}
\def\g{\gamma}
\def\d{\delta}
\def\e{\epsilon}
\def\se{\varepsilon}
\def\E{\Sigma}
\def\t{\theta}
\def\k{\kappa}
\def\l{\lambda}
\def\t{\theta}
\def\s{\sigma}
\def\r{\rho}
\def\p{\psi}
\def\f{\phi}
\def\F{\Phi}
\def\vf{\varphi}
\def\x{\xi}
\def\G{\Gamma}
\def\adot{\dot{\alpha}}
\def\bdot{\dot{\beta}}
\def\gdot{\dot{\gamma}}
\def\ddot{\dot{\delta}}
\def\cbar{\bar{c}}
\def\fbar{\bar{\phi}}
\def\vfbar{\bar{\varphi}}
\def\lbar{\bar{\lambda}}
\def\sbar{\bar{\sigma}}
\def\pbar{\bar{\psi}}
\def\xbar{\bar{\xi}}
\def\tbar{\bar{\theta}}
%
\def\ds{\displaystyle}
\def\be{\begin{equation}}
\def\ee{\end{equation}}
\def\bea{\begin{eqnarray}}
\def\eea{\end{eqnarray}}
\def\ov{\overline}
%

\thispagestyle{empty}
\begin{center}
{\Large{\bf N=1 SYM Action and BRST Cohomology }}
\vskip1.5em
K. \"{U}lker\\\vskip3em
{\sl Istanbul Technical University}\\
{\sl Science and Letters Faculty Physics Dept.}\\
{\sl 80626,  Maslak - Istanbul, Turkey}\\ 
\vskip1.5em
\end{center}
\vskip2em
%
\begin{abstract}
The relation between BRST cohomology and the N=1 supersymmetric Yang-Mills action in 4 dimensions is discussed. In particular, it is shown that both off and on-shell N=1 SYM actions are related to a lower dimensional field polynomial by solving the descent equations, which is obtained from the cohomological analysis of linearized Slavnov-Taylor operator $\B$, in the framework of Algebraic Renormalization. Furthermore we show that off and on-shell solutions differ only by a $\B$- exact term, which is a consequence of the fact that the cohomology of both cases  are same.
\end{abstract}
\vskip5em
PACS codes: 12.60.Jv, 11.15.-q

\vskip4em
{\small{e-mail: kulker@itu.edu.tr}}
\vfill
\eject
\setcounter{page}{1}
%
%
\section{Introduction}

Supersymmetric versions of renormalizable quantum field theories are famous for their finiteness properties due to the cancellations of ultraviolet divergences \cite{zumino} and since then these cancellations are verified by using the superfield formalism in the superspace that displays all the simplifications due to supersymmetry for various models \cite{fujikawa}. 

In the component field formalism these non-renormalization theorems are first derived in Wess-Zumino model by representing the interaction terms as a multiple supervariation of a lower dimensional field monomial \cite{flume} and a similar analysis is extended in Ref.\cite{kraus1} and \cite{kraus2} to SQED and N=1 SYM cases respectively. These results show that the cohomological structure of the supersymmetric models can be thought as the algebraic source of the non-renormalization theorems, and therefore it is worthwhile to discuss this structure not only for its own interest but also for its use in proving the non-renormalization theorems.

When the cohomological structure of supersymmetric actions and interaction terms are studied it is found that to construct these terms from super-variations of lower dimensional field polynomials it is crucial to use the supersymmetry algebra with auxiliary fields \cite{ulker} \footnote{For a similar approach of constructing N=1 globally and locally supersymmetric actions and also for the discussion of anomalies, see \cite{bra2}.}. In other words, if one begins with supersymmetric theories without auxiliary fields (the theories with on-shell supersymmetry), one encounters problems due to the fact that supersymmetry algebra realized without auxiliary fields closes modulo equation of motion terms. For instance when super Yang-Mills actions are intended to be derived by using pure on-shell supervariations of lower dimensional field monomials the resulting expressions differ from the original one (that can be found by off shell variations) by equation of motion terms \cite{ulker}. There is not a non-trivial way to restore these missing terms.

An elegant way to overcome the problem of above mentioned on-shell closure of the algebra is to use Batalin-Vilkovisky formalism \cite{batalin} that is often called algebraic renormalization \cite{psbook} where the anti-fields of the corresponding fields are introduced to control the renormalizability of a theory. It is well known that in this formalism the renormalization program reduces to an algebraic discussion of the cohomology of a linearized Slavnov-Taylor operator in the space of local field polynomials. The counter terms and the possible anomalies are then the solutions of this cohomology problem with ghost number 0 and 1 respectively \cite{psbook}. In this framework renormalization of supersymmetric theories has been discussed for various supersymmetries \cite{w,mag,mpw} by extending the BRST transformations to include supersymmetry transformations \footnote{Such an extension of BRST transformations that includes rigid symmetries is first introduced in \cite{bon}. The problem of how to extend the BRS formalism to include arbitrary global symmetries can be found in Ref. \cite{bra1}.} and it is shown that the on-shell closure of the algebra can be rectified by adding some (none standard) quadratic terms in the anti-fields of fermions to the action \cite{w,mag,mpw}. (See also \cite{ps})

In this paper our aim is to show that both off and on-shell supersymmetric (extended) action of 4 dimensional N=1 super Yang-Mills theory can be related to the lower dimensional field monomial $tr\l\l$ by studying afore mentioned (extended) BRST cohomology in the Algebraic Renormalization framework. 

Our strategy and results are as follows. After a brief description of the theory and introducing the corresponding linearized Slavnov-Taylor operator $\B$, we study the solutions of descent equations which arise from the cohomological analysis of $\B$. In particular, by solving these equations for ghost number 0, we derive algebraic identities that relate the 3-dimensional gauge invariant field monomial $tr\l\l$ to both off and on-shell supersymmetric (extended) Yang-Mills actions. It is then found that these two solutions for off and on-shell supersymmetric cases differ from each other only by a $\B$-exact term which shows that the BRST cohomology of both cases are the same. The missing terms that are found by climbing up with pure on-shell supervariations in Ref.\cite{ulker} are then restored with the help of this $\B$-exact term.

%
%
\section{N=1 SYM Theory}
We begin our analysis with the off-shell supersymmetric Yang-Mills theory in Wess-Zumino gauge. The action,
\be
S_{SYM} = \frac{1}{g^2} tr \int d^4 x \lb - \frac{1}{4}F^{\mu\nu} F_{\mu\nu}  - i\l \Ds \lbar +\yarim \D^2 \rb  
\ee
contains the N=1 vector multiplet $(A_\mu , \l , \lbar ,\D)$ whose components transform under the adjoint representation of the gauge group. \footnote{Throughout this paper we use Wess and Bagger conventions \cite{wbbook}.} The action is invariant under the following supersymmetry transformations:
\be 
\d A_{\mu}=i{\t}{\s} _{\mu} \bar{{\l}}+ i\bar{\t}\bar{\s} _{\mu}{\l}
\ee
\be
\d{\l}={\s}^{\mu\nu}{\t}F_{\mu\nu}+ i\t \D  \qquad, \qquad 
\d\D= -\t D\!\!\!\!/\bar{\l} +\tbar \bar{D}\!\!\!\!/ \l
\ee
where
\be
\d = \t^\a \qu _{\a} + \tbar_{ \adot} \qubar ^{\adot}
\ee
and \(\t, \tbar \) are the anti-commuting supersymmetry parameters. The algebra reads
\be 
\{ \qu  ,\qu  \}  =\, 0 \, =  \{ \qubar ,\qubar \}  \quad ,\quad \{ \qu  ,\qubar \}   = \, -2i\s^{\mu} D_{\mu}  . 
\ee

Following the standard procedure \cite{w,mag,mpw,psbook,bra1}, the extended BRST generator $s$, that contains the supersymmetry transformations, can be defined on the vector multiplet as  
\be
s := s_0 -i\x^\a \qu _\a -i \xbar_{\adot} \qubar ^{\adot}
\ee
where $s_0$ is the ordinary BRST generator and since $s_0$ is anti-commuting $\x , \xbar$ are taken to be commuting constant spinor parameters. On the other hand, $s_0$ carries ghost number and therefore, $\x , \xbar$ can be promoted to the status of global ghosts that corresponds to the rigid supersymmetry.  

The action of $s$ on the vector multiplet reads:
\bea
sA_\mu &=& D_\mu c + \x \s_\mu \lbar + \xbar \sbar_\mu \l \\
s\l &=& i\{c,\l \} - i\s^{\mu\nu} \x F_{\mu\nu} +\x\D \\
s\lbar &=& i\{c,\lbar \} - i\sbar^{\mu\nu} \xbar F_{\mu\nu} -\xbar \D  \\
s\D &=& i[c, \D] +i\x \Ds \lbar -i\xbar \Dsbar \l \\
sc &=& \frac{i}{2}\{ c,c \} - 2i\x \s^\nu \xbar A_\nu    .
\eea
Note that Faddeev-Popov ghost field $c$ is asked to transform under supersymmetry to close $s^2$ on translations \footnote{We could add the generator of translations inside $s$ but since our aim is to work with the integrated polynomials our definition will cause no problems.}:
\be
s^2 = -2i \x \s^\nu \xbar \del_\nu
\ee
The action (1) is obviously invariant under $s$ since it has both gauge and supersymmetry invariance. The gauge freedom of the action is fixed in the usual way by adding a trivial co-cycle\footnote{Note that in the standard BV formalism, anti-fields of the trivial pair are also introduced i.e. $\Sigma^A=\{\vf^a,\cbar,b\}$, $\Sigma_A ^*=\{\vf_a ^*,\cbar^*,b^*\}$ where $\vf^a=\{A_{\mu},\l,\lbar,\D,\c\}$ so that $I=S_{SYM}+tr\int d^4x \Sigma_A ^*s\Sigma^A$. Then the gauge fixing is done with the help of a gauge fixing fermion $\Psi$ by setting $\Sigma_A ^*=\frac{\d\Psi}{\d\Sigma^A}$ which gives a gauge fixing term $S_{gf}$. However, since $\frac{\d\Psi}{\d\Sigma^A}s\Sigma^A=s\Psi$, this is equivalent to adding a s-exact term to the orginal action (1). See for instance \cite{weinberg}.

In our work, gauge fixing term (13) corresponds to the choice $\Psi =-tr\int d^4 x \cbar\del_{\mu}A^{\mu}$ (Landau gauge). } 
\be
S_{gf} = -tr \int d^4 x s(\cbar \del^\mu A_\mu)
\ee
with the help of the trivial pair $(\cbar, b)$
\be
s\cbar = b \quad ,\quad sb= -2i\x \s^\nu \xbar \del_\nu \cbar
\ee
The action is further extended by introducing the anti-fields, that couple to the s-transformations of the corresponding fields, in order to be able to write Slavnov-Taylor (ST) identity :
\be
S_{ext} = tr\int d^4x (A^* _\mu sA^\mu + \l^* s\l + \lbar^* s\lbar + \D^* s\D + c^* s c)
\ee
The total action is now given by,
\be
I = S_{SYM} + S_{gf} + S_{ext}
\ee
and satisfies the following ST identity,
\be
\S(I)=  2i \x \s^\nu \xbar \Delta_\nu
\ee
where
\bea
\S(I)&=&tr \int d^4 x ( \frac{\d I}{\d A_\mu ^*}\frac{\d I}{\d A^\mu} + \frac{\d I} {\d \l^{*\a} }\frac{\d I}{\d \l_\a} + \frac{\d I}{\d \lbar _{\adot}^*} \frac{\d I}{\d \lbar ^{\adot} } 
+ \frac{\d I}{\d \D^*}\frac{\d I}{\d \D}  \nonumber \\
&&\qquad \qquad +\frac{\d I}{\d c^*}\frac{\d I}{\d c} + s\cbar \frac{\del I}{\del \cbar} + sb \frac{\del I}{\del b})
\eea
and $\Delta_\nu$ is a classical breaking due to (12),
\be
\Delta_\nu = tr\int d^4 x (A_\mu ^* \del_\nu A^\mu - \l^* \del_\nu \l - \lbar^*  \del_\nu \lbar + \D^* \del_\nu \D - c^* \del_\nu c )    .
\ee
that has no effect on the renormalization of the theory \cite{psbook}.

Note that, if the gauge invariance and supersymmetry was treated separately one would need an infinite number of sources (antifields) \cite{breit}. As a matter of fact of collecting all the symmetries in a unique BRST generator with the help of global ghosts, one avoids such an infinite number of  sources. Then the Ward Identities are transformed into a Slavnov-Taylor identity that includes all these symmetries and Eq.(17) can be used to analyse the renormalization of supersymmetric gauge theories \cite{w,mag,mpw}.   

The relevant object for cohomological analysis is the linearized ST
 operator,
\bea
\B_I&=&tr\int d^4x  ( \frac{\d I}{\d A_\mu ^*}\frac{\d }{\d A^\mu} +  \frac{\d I}{\d A_\mu}\frac{\d }{\d A^{*\mu}} \nonumber \\ 
&&\qquad + \frac{\d I} {\d \l _\a }\frac{\d }{\d \l ^{*\a}} + \frac{\d I} {\d \l ^{*\a} }\frac{\d }{\d \l_{\a}} +\frac{\d I}{\d \lbar ^{\adot}} \frac{\d }{\d \lbar _{\adot}^* } + \frac{\d I}{\d  \lbar _{\adot}^*  } \frac{\d }{\d \lbar ^{\adot} } \\
  &&\qquad+\frac{\d I}{\d \D}\frac{\d }{\d \D^*}  +\frac{\d I}{\d \D^*}\frac{\d }{\d \D}+ \frac{\d I}{\d c^*}\frac{\d }{\d c} + \frac{\d I}{\d c}\frac{\d }{\d c^*}  +s\cbar \frac{\del }{\del \cbar} + sb \frac{\del }{\del b} )\nonumber
\eea
which reads
\be
\B_I \f  = s\f  \quad ,\quad \B_I \f^* = \frac{\d I}{\d \f}
\ee
where $\f$ denotes collectively all the fields and $\f^*$ the corresponding anti-fields.

The crucial property of $\B_I$ is that it satisfies the following algebraic relation that is derived with the help of eq.(17):
\be
\B_I \B_I = -2i\x \s^\nu \xbar \P_\nu 
\ee
\be
\P_\nu = \sum tr \int d^4 x ( \del_\nu \f \frac{\del}{\del \f} + \del_\nu \f^* \frac{\del}{\del \f^*})\nonumber 
\ee

The on-shell case is obtained , as usual, by eliminating the auxiliary field $\D$ with its equation of motion, that is $\D=0$ for pure SYM theory. The extended BRST transformation $\tilde{s}$ is given now by 
\be
\tilde{s} = s \vert _{\D=0}  
\ee
and since, for on-shell case supersymmetry algebra (5) holds only modulo equations of motion of fermions, $\tilde{s} ^2$ also closes on translations modulo equations of motion 
\be
\tilde{s}^2 = -2i \x \s^\nu \xbar \del_\nu \qquad(modulo \, eq.\, of\, motion \, of \, \l , \lbar) .
\ee
However, it is still possible to have a Slavnov-Taylor operator that also  squares to a boundary term by adding a quadratic term in the anti-fields to the extended action (16). For N=1 SYM it is found to be:
\be
S_{quad}= - \yarim g^2tr \int d^4  x(\x\l^* - \xbar\lbar ^*)^2
\ee
The total classical action now reads,
\be
\tilde{I} = S_{SYM} \vert _{D=0}+S_{gf}+S_{ext}\vert _{\D=0}+S_{quad} 
\ee
and as a consequence, relations (18) and (20) take the form. 
\be
\S( \tilde{I} )=  2i \x \s^\nu \xbar \tilde{\Delta} _\nu
\ee
\be
\B_{\tilde{I}} \B_{\tilde{I}} = -2i\x \s^\nu \xbar \tilde{\P} _\nu 
\ee
for $\tilde{\Delta} _\nu =\Delta_\nu \vert_{\D=0}  \, ,\,  \tilde{\P} _\nu =\P_\nu \vert_{\D=0}$.
This fact is not surprising since the combination $g^2(\x \l^* - \xbar \lbar^*)$ exactly behaves like the auxiliary field $\D$ i.e.,
\be
\B_{\tilde{I}} \l = \tilde{s} \l + g^2\x (\x \l^* - \xbar \lbar^*) \quad ,\quad \B_{\tilde{I}} \lbar = \tilde{s} \lbar - g^2\xbar (\x \l^* - \xbar \lbar^*) 
\ee
\be
\B_{\tilde{I}}g^2 (\x \l^* - \xbar \lbar^*) = B_I \D \vert_{\D=g^2 (\x \l^* - \xbar \lbar^*)}
\ee

%
%
\section{The Construction of the Action from the Cohomology of ST Operator}

The starting point of analysing the cohomology of the operator \footnote{For the following discussion, $\B$ will stand for both $\B_I$ and $\B_{\tilde{I}}$.} $\B$  is to note that when its action is restricted on the space of integrated polynomials, the r.h.s. of equations (22) and (29) become a total derivative. Therefore, if one assumes that there is no boundary contribution, $\B $ becomes a nilpotent operator on this space and therefore the equation 
\be
\B  \int d^4 x \X = 0
\ee
constitutes a cohomology problem on the space of integrated local polynomials of fields and anti-fields. It is well known that the non-trivial solutions (that are not a trivial co-cycle, $\X \ne \B \X'$) with dimension four, vanishing R-charge and the ghost number 0 and 1 are directly related with the counterterms and the anomalies respectively, in the Algebraic Renormalization method \cite{psbook}.  

We can derive the descent equations by taking the local version of equation (32): 
\be
\B \X_0 = (\xbar \sbar _\mu )^\a \del ^\mu \X_{1\a}
\ee
The $(\xbar \sbar _\mu )$ factor in front of the derivative is assumed to appear due to supersymmetry algebra. The second of descent equations is found by applying $\B$ to equation (33) and using the local version of equations (22) and (29). It reads 
\be
\B \X_{1\a} = -2i \x_\a \X_0 + (\xbar \sbar _\mu )^\b \del ^\mu \X_{2\a \b}
\ee 
where here it is crucial that $\X_{2\a \b}$ is anti-symmetric in its spinor indices. Similarly taking a further $\B$ variation of equation (34) third of descent equations is obtained as:
\be
\B \X_{2\a\b} = 2i \x_\a \X_{1\b} -2i \x_\b \X_{1\a}
\ee 
Descent equations terminate at this level since a rank 3 anti-symmetric tensor in spinor indices is zero identically. 
 
Solutions of the lowest descent equation (35) can be found by using the symmetry content of the theory. For this purpose we summarize the dimension, ghost number and the R-charges of the fields, anti-fields, ghosts and the operator $\B$ in table 1.
%
\begin{table}[hbt]
\centering
\begin{tabular}{|c||c|c|c|c|c|c|c|c|c|c|c|c|}
\hline
&$A_\mu$&$\l$&$\D$&$c$&$\cbar$&$b$&$\x$&$A^*_\mu$&$\l^*$&$\D^*$&$c^*$&$\B$ \\
\hline
$d$&1&3/2&2&0&2&2&-1/2&3&5/2&2&4&0 \\
\hline
$GP$&0&1&0&1&1&0&0&1&0&1&0&1 \\
\hline
$Gh$&0&0&0&1&-1&0&1&-1&-1&-1&-2&1\\
\hline
$R$&0&-1&0&0&0&0&-1&0&1&0&0&0 \\
\hline
\end{tabular}
\caption[t1]{Dimensions   $d$, Grassmann parity $GP$, ghost number $Gh$
  and R-weights.}
\end{table}

Since our aim is to show that the N=1 SYM action belongs to the cohomology of $\B$ we will concentrate on a gauge invariant solution $\X_0$ that has dimension 4 with vanishing ghost number and R-charge and Grassmann even. Therefore a solution $\X_2$ of eq.(35) have dimension 3, R-charge -2. The only field polynomial satisfying these conditions is $tr(\l\l)$ and the solution with desired index structure to the descent equation (35) is obtained as \be
\X_{2\a \b}=k \e_{\a\b} tr(\l\l)
\ee
for both off and on-shell cases where k is a constant. Here it is important to note that the solution of the lowest descent equation (35) is proportional to the lowest component of the multiplet that the action belongs. This is related to the fact that the number of descent equations depends on the structure of global ghosts $\x_{\a},\xbar_{\adot}$ and therefore to supersymmetry.

The solution $\X_1$ can be directly read from the r.h.s. of equation (35) after substituting (36). The off-shell solution is found to be
\be
\X_{1\a}^{off} = ik tr(-i \s^{\mu\nu\, \b}_\a \l_\b F_{\mu\nu} - \l _\a \D) 
\ee
and the on-shell one as
\be
\X_{1\a}^{on} = ik tr\lb -i \s^{\mu\nu\, \b}_\a \l_\b F_{\mu\nu} -g^2(\yarim \l _\a \x \l^* + \yarim \l _\a ^* \x \l - \l_\a \xbar \lbar ^*) \rb 
\ee
However, due to the commuting nature of $\x$ adding a term  proportional to $\x_\a$ with correct quantum numbers and dimension to $\X_{1\a}$ will not affect the l.h.s. of (35). We choose this term to be 
\be
\X_{1\a}^t = \frac{i}{2} g^2 k \x_\a tr \l^* \l
\ee
in order to have a similar solution with $\X_{1\a}^{off}$ when $\D$ and $g^2 (\x \l^* - \xbar \lbar ^*)$ are interchanged:
\be
\tilde{\X}_{1\a}^{on}= \X_{1\a}^{on}+\X_{1\a}^t= ik tr\lb -i \s^{\mu\nu\, \b}_\a \l_\b F_{\mu\nu} - \l_\a g^2(\x\l^* -\xbar\lbar^*)\rb
\ee

Finally by substituting $\X_{1\a}^{off}$ and $\tilde{\X}_{1\a}^{on}$
to equation (34) we end up with the desired results for both off and on-shell supersymmetric cases:
\be
\X_0 ^{off} = k tr\lb - \frac{1}{4}F^{\mu\nu} F_{\mu\nu} -\frac{i}{8} \epsilon ^{\mu\nu\l\k}F_{\mu\nu}F_{\l\k} - i\l \Ds \lbar +\yarim \D^2 \rb
\ee
\be
\X_0 ^{on}=k tr\lb - \frac{1}{4}F^{\mu\nu} F_{\mu\nu} -\frac{i}{8} \epsilon ^{\mu\nu\l\k}F_{\mu\nu}F_{\l\k} - i\l \Ds \lbar + \yarim g^4 (\x\l^* -\xbar \lbar^*)^2\rb 
\ee
and since the topological term $-\frac{i}{8} \epsilon ^{\mu\nu\l\k}F_{\mu\nu}F_{\l\k}$ can be written as a total derivative we get the following relations between the solutions of equation (32) and the extended actions (16) and (27)   
\be
\int d^4 x \X_0^{off} = k \frac{d}{dg^{-2}} I  \qquad , \qquad \int d^4 x \X_0^{on} = k \frac{d}{dg^{-2}} \tilde{I} 
\ee
from which we conclude that $\frac{dI}{dg^{-2}}$ and $\frac{d\tilde{I}}{dg^{-2}}$  are the invariant counterterms and that they belong to the cohomology of $\B_I$ and $\B_{\tilde{I}}$, in the space of local functionals depending on the coupling constant $g$.\footnote{The arbitrary coefficient $k$ here, can be thought as a parameter corresponding to a possible renormalization of the coupling constant $g$.} This result is in agreement with that of Ref.\cite{mpw}.

On the other hand, above solutions (36-42) of the descent equations can be related to each other with a climbing up operator. For this purpose we filter the operator $\B$ with respect to supersymmetry ghost $\x$ as
\be
\mathcal{N} = \x^\a \frac{\d}{\d \x^\a}\quad ;\quad \B = \sum{\B_n} \quad ,\quad [\mathcal{N}, \B_n] = n \B_n
\ee 
which leads to the algebra 
\bea
\B_0 ^2 &=&0\\
\{\B_0,\B_1\}&=&-2i\x \s^\mu \xbar \del_{\mu}\\
\{\B_0,\B_2\}+\B_1 ^2 &=&0\\
\B_2 ^2 = \{\B_1,\B_2\}&=&0
\eea
We further define the operators $\b$ and $\q_\a$ with respect to the following expansion of $\B$ \footnote{Obviously $Z^{on} \not= Z^{off}$ and in the case of off-shell supersymmetry $X_{\a\b}^{off} = 0 = \bar{X}_{\adot\bdot} ^{off}$.}

\be
\B = \d + \xbar _{\adot} \bar{Y} ^{\adot} + \x ^\a Y _\a 
+ \xbar_{\adot} \x^\a Z^{\adot} _\a + \yarim \xbar _{\adot} \xbar _{\bdot} \bar{X} ^{\adot \bdot}+ \yarim \x^\a \x^\b X_{\a \b}
\ee  
\be
\b = \B_0 = \d + \xbar _{\adot} \bar{Y} ^{\adot}+ \yarim \xbar _{\adot} \xbar _{\bdot} \bar{X} ^{\adot \bdot}
\ee
\be
\x^\a \q_\a = \B_1 + \B_2=\x^\a ( Y _\a 
+ \xbar_{\adot}Z^{\adot} _\a+ \yarim \x^\b X_{\a \b})  
\ee
By using the above algebra and definitions, we see that the lowest descent equation separates into two, 
\be
\b \X_{2\a \b} =0
\ee
and 
\be
\x^\g \q_\g \X_{2\a\b} = 2i \x_\a \X_{1\b} -2i \x_\b \X_{1\a}
\ee 
Note that the equation (51) indicates that the operator $\q_\a$ can be used as a kind of climbing up operator. Indeed, after some algebra it is found that the solutions of descent equations (33-35) are algebraically related to each other as 
\be
\q_\g  \X_{2\a\b} = 2i( \e_{\a\g} \X_{1\b} -\e_{\b\g} \X_{1\a})
\ee
\be
\q_\a \X_{1\b} = 2i \e_{\a\b}\X_0
\ee
and we get the following lift from the bottom to the top: 
\be
\q^\a \q^\b \X_{2\a\b} = -8 \X_0
\ee

Therefore off and on-shell extended actions can be related to the lower dimensional field monomial $tr \l\l$ as a consequence of the relation (56) for $k=\frac{1}{g^2}$: 
\bea
-\yarim g \frac{d}{dg} I &=&  S_{SYM}\nonumber \\ 
&=& - \frac{1}{8g^2} \q^\a \q_\a tr \int d^4 x\l^\b \l_\b 
\eea
\bea
 -\yarim g \frac{d}{dg} \tilde{I} &=& S_{SYM}\vert _{\D =0} - S_{quad}
\nonumber \\
&=& - \frac{1}{8g^2} \q^\a \q_\a tr\l^\b \l_\b - \frac{1}{4}\B_{\tilde{I}} tr\int d^4 x\l^{*\a}\ \l_\a \qquad .
\eea
which shows that both off and on-shell supersymmetric SYM actions can be constructed from 3 dimensional field monomial $tr \l\l$.

The difference between equation (57) and (58) is a $\B$ exact term. This result can be related to the theorem that local BRST cohomologies of two formulations of the same theory differing in auxiliary field content are same \cite{bbh}. Moreover, note that climbing up from $tr\l\l$ with the chiral (or anti-chiral) part of ordinary on-shell supersymmetry transformations gives an action modulo equation of motion term \cite{ulker}. Here this missing term is restored by adding a $\B$-exact term including the anti-field $\l^*$ since a $\B$ transformation of an anti-field includes the equation of motion of the corresponding field in a well defined way due to eq.(21) \footnote{This part of the operator $\B$ is often called in the literature Koszul-Tate differential. See Ref. \cite{bbhrev} for its importance for BRST cohomological calculations.}.

\section{Conclusion}

In conclusion, we have seen that the relations (57) and (58) imply the above method of working the BRST cohomology through the descent equations, gives the relation between the action and the lowest component of the multiplet that the action belongs in an elegant way for both off and on-shell supersymmetric cases. This method can also be extended for the N=2 SYM theory and we believe that it can provide a natural and simpler way of proving the non-renormalization theorems of various supersymmetric Yang-Mills theories.\footnote{Work in progress. See also \cite{soretal} for a similar approach to the twisted version of N=2 (on-shell) supersymmetric Yang Mills theory, where the non-renormalization of the coupling constant beyond one loop has been shown.}.
   
Finally, it is worthwhile to remark that for on-shell supersymmetric case certain combination of anti-fields, here $g^2(\x \l^* - \xbar \lbar^*)$, behaves exactly as the auxiliary field for the solutions of all levels of descent equations. (Note also the sign of $S_{quad}$ in eq.s (27) and (42,58).) This observation also holds for the N=2 case. It can be interesting to find out if this fact can be used to obtain an off shell formulation of the supersymmetric theories where the auxiliary field content is not known, such as N=4 SYM theory.

%
%
%
%
{\bf Acknowledgements: }
The author is grateful to R. Flume for introducing the subject. He also gratefully acknowledges the enlightening discussions with \"{O}. F. Day\i{} and M. Horta\c{c}su.\\This work is supported by TUBITAK, Scientific and Technical Research Council of Turkey, under BAYG-BDP.

\vskip3em

%
%
%
%

\end{document}